В.В. Мигунов

# СРЕДА РАЗРАБОТКИ ПРОГРАММ ПАРАМЕТРИЧЕСКОЙ ГЕНЕРАЦИИ ЧЕРТЕЖЕЙ В САПР РЕКОНСТРУКЦИИ ПРЕДПРИЯТИЙ

Известно, что по мере развития практически любая САПР приходит к необходимости использования языка разработки приложений. AutoCAD 2004 имеет даже 4 интерфейса прикладного программирования: Visual LISP, VBA, ActiveX и ObjectARX. Отечественная СПРУТ-технология [1] является, по сути, средой разработки и исполнения приложений (силами самих конструкторов и технологов), графические работы в ней носят вспомогательный характер. В САПР реконструкции предприятий роль среды разработки приложений не может быть столь обширной по двум причинам. Результатом работ проектировщика в этом случае является чертеж для выполнения строительно-монтажных работ, в которых электронное представление объекта практически не может использоваться. Объекты реконструкции, задающие дополнительные "начальные и краевые" условия работы приложений по сравнению с проектированием нового изделия, отличаются значительной индивидуальностью и, если и имеют электронное представление, то лишь в форме чертежей, но не алгоритмов их синтеза. Тем не менее, есть ряд типовых проектных решений где удобно применить программы параметрической генерации чертежей (ППГ). Ниже описаны основные идеи, лежащие в основе среды разработки таких программ, реализованной в САПР TechnoCAD GlassX [2].

С точки зрения пользователя среда включает процедуры редактирования и компиляции исходных текстов ППГ, процедуры отладочного исполнения откомпилированных программ, процедуры ведения библиотек ППГ и обеспечивает:
- удобный выбор ППГ - прототипа из имеющихся в библиотеках;
- диверсифицированный контроль корректности исходных текстов ППГ;
- загрузку редактора текстов, в том числе с выдачей сообщения об ошибке с указанием ее положения в исходном тексте ППГ курсором;
- компиляцию исходного текста ППГ в компактный исполнимый код;
- запись на диск и просмотр протокола компиляции;
- исполнение ППГ и просмотр результатов ее работы;
- занесение откомпилированных ППГ в библиотеки с комментариями;
- выдачу контекстной справочной информации.

Основные элементы языка ППГ соответствуют по смыслу элементам Pascal (типы данных, переменные, строки…) и FoxPro (endif, endcase…). Надежность работы программ повышается за счет снижения быстродействия. Исключены средства непосредственной работы с оперативной памятью, ужесточен контроль типов данных, исключена вариантность вызовов функций (нефиксированное число параметров и неопределенность их типов), ужесточены требования к структурным разделам программ, к записи всех утверждений языка.

ППГ включают встроенные типы данных и операции, аналогичные имеющимся в TechnoCAD GlassX: геометрические элементы, функции организации пользовательского интерфейса, изменения установок, добавления в чертеж и др. Состав операций допускает расширение и отвечает текущим потребностям решаемых задач.

Ниже приведена часть встроенных типов данных с использованием обозначений для записей и массивов, принятых в Pascal. Базовые типы: Логическое, Целое, Вещественное, Строка, Адрес (служебный, пользователь с ним не работает).
- Длина .R: Вещественное
- Точка .X: Вещественное; .Y: Вещественное
- Отрезок .Начало: Точка; .Конец: Точка
- Окружность .Центр: Точка; .R: Вещественное
- Дуга .Окр_ть: Окружность; .Угол1: Вещественное; .Угол2: Вещественное

- <u>Массив_углов</u> : Массив [0..15] из "Точка"
- <u>Ломаная</u> .Nотр: Целое; .Углы: Массив_углов
- <u>Текст</u> .Сноска: Точка; .ЛучТекста: Луч; ._АдрТекста: Адрес
- <u>Линейный_размер</u> .База: Отрезок; .Начало: Точка; .ЛучТекста: Луч; .Текст: Строка;
- <u>Атрибут</u> .Слой: Целое; .Цвет: Целое; .Тип_Линии: Целое; .Сист_Отсчета: Целое

Допускается также объявление пользовательских записей и массивов.

В ППГ используются встроенные константы: число π Pi; логические Да, Нет; цвета Черный, Синий…Желтый, Белый; единицы измерения длин: Натура, Бумага; типы линий: Сплош_осн, Сплош_тонк, Штрих_утол, Штриховая, Пункт_тонк, Пункт_утол, Разомкнутая.

Все действия ППГ называются операциями, включая арифметические, управляющие и др. Операции, смысл которых очевиден из наименования: *, + (для чисел и строк), –, /, DIV, MOD, INT, FRAC, ROUND, ABS, ^, SQRT, LN, EXP, LG, IIF, ИзГрадВРад, ИзРадВГрад, ЧислоВСтроку, СтрокаВЦелое, Подстрока, СтрокаВЧисло, <, <=, <>, =, >, >=, NOT, AND, OR, XOR, SIN, COS, TG, ARCSIN, ARCCOS, ARCTG, SH, CH, TH, ARSH, ARCH, ARTH. Присвоение значения, в том числе пользовательским массивам и записям, обозначается ":=". В таблице приведены примеры других операций.

| Наименование | Назначение |
|---|---|
| *Управляющие операции* | |
| GOTO | безусловный переход на метку |
| EXIT | останов выполнения |
| IF | условный переход |
| ELSE | метка "иначе" условного перехода |
| ENDIF | конец команд, обходимых в условном переходе |
| CASE | множественный условный переход |
| ON | условие множественного условного перехода |
| ONELSE | метка "иначе" множественного условного перехода |
| ENDCASE | конец команд, обходимых в множественном условном переходе |
| *Организация диалога с пользователем* | |
| Сообщение | выдача сообщения в середине экрана |
| Информация | выдача сообщения в верхней информационной строке |
| Запрос | запрос ответа Да, Нет или Отказ |
| НовоеМеню | создание меню |
| ДобОпцию | добавление опции в меню |
| Доб_5_Опций | добавление в меню от 1 до 5 опций |
| ПоказМеню | запрос выбора в меню |
| МенюИзФайла | запрос выбора в меню, сформированного из текстового файла |
| ТекстОпции | возвращает текст выбранной опции меню |
| Новая_форма | создание формы ввода |
| Новое_поле | добавление поля в форму ввода |
| Новое_полеXY | добавление поля в форму ввода в указанных координатах |
| Масштаб_поле | добавление в форму ввода поля в указанных координатах с возможностью выбора масштаба из меню |
| Редактор | запрос ввода/редактирования данных в форме ввода |
| *Работа с элементами чертежа TechnoCAD GlassX* | |
| Глоб_Атр | вернет текущие глобальные установки атрибутов TechnoCAD GlassX |
| Уст_Атр | установка атрибутов – слой, цвет, тип линии, система отсчета |
| Отрез | добавление в чертеж отрезка |
| Прямоуг | добавление в чертеж прямоугольника |
| ДугаОкружн | добавление в чертеж дуги окружности |
| ЛРазмСноски | изменение глобальной установки сносок линейного размера |

| Наименование | Назначение |
|---|---|
| ЛРазмТочн | изменение глобальной установки десятичной точности лин. размера |
| ЛРазмВынос | изменение глобальной установки выносных линий линейного размера |
| ЛРазмШрифт | изменение глобальной установки шрифта линейного размера |
| ЛРазмСтрелки | изменение глобальной установки наконечников линейного размера |
| ГорРазмер1 | добавление в чертеж простого горизонтального линейного размера |
| ВерРазмер1 | добавление в чертеж простого ветикального линейного размера |
| РамкаРазм | добавление в чертеж двух размеров, образмеривающих прямоугольник |
| ТекстСноска | изменение глобальной установки сноски текста |
| ТекстШрифт | изменение глобальной установки шрифта и шага строк текста |
| ДлинаСтроки | вернет длину строки в мм при текущих глобальных установках текста |
| НачатьТекст | инициализация текста (задается первая строка) |
| ДобСтроку | добавление строки в многострочный текст |
| ОтмВысоты | добавление в чертеж отметки высоты |
| ОбрывТрубы | добавление в чертеж изображения обрыва трубы или арматуры |
| ОбрывПоДуге | добавление в чертеж изображения дуговой линии обрыва |
| УбратьИзЧерт | удаление из чертежа элемента с заданным номером |

Текст ППГ состоит в общем случае из трех разделов:
PROGRAM <Имя программы>
[TYPE..ENDTYPE] - объявления пользовательских типов (необязательный раздел)
[VAR ..ENDVAR] - объявления переменных (необязательный раздел)
  исполняемая часть (обязательный раздел)
ENDROGRAM

В исполняемой части циклы for, while, repeat и операторные скобки begin end не поддерживаются, вместо них используются goto и метки, endif, endcase. В целях краткости и ясности текста ППГ лишние пары скобок считаются ошибкой. Имена переменных, меток и операций разрешены длиной до 30 символов из набора [a..z, A..Z, а..я, А..Я, 0..9, _]. Поля записей отделяются точкой.

В примере простой ППГ ниже для экономии места часть строки слиты, но в тексте ППГ соблюдается правило "одна строка - одно утверждение",.каждая строка заканчивается ";".
   program Оголовок вентпанелей;
  var;
   Вид, НомерЭл : Целое;   { Выбранный из меню вид, номер элемента в чертеже }
   H, L, B, n1, n2, n3, n4 : Вещественное; { Габариты разреза и вспомогательные числа }
   Шапка : Атрибут;   { Шапка элемента для включения в чертеж }
  endvar;
   Шапка := Глоб_Атр; Шапка.Цвет := Черный; Шапка.Тип_Линии := Сплош_осн;
   Шапка.Сист_Отсчета := Натура;
   ЛРазмТочн (0);          { Меняет установки точности линейных размеров }
   ЛРазмВынос (1, 1, 1.5);   { Меняет установки выносных линейных размеров }
   ЛРазмШрифт (3.5, 0, 0.8); { Меняет установки шрифта в линейных размерах }
   ЛРазмСтрелки (3, SQRT (2), 3, SQRT (2)); {Меняет установки стрелок в LSize}
   L := 880; B := 450; H := 600;
   НовоеМеню ('Оголовок вентпанелей'); ДобОпцию (' Вид сверху', 1, Да);
   ДобОпцию (' Вид спереди ', 2, Да);      ДобОпцию (' Вид сбоку', 3, Да);
   Вид := ПоказМеню (1); if Вид = 0; exit; endif;
   case;
    on Вид = 1; { Вид сверху, три отверстия }
     НомерЭл := Прямоуг (Шапка, 0, 0, L, B);
     n1 := B / 2; n2 := L / 6; n3 := n2 / 2; n4 := n1 / 2;
     НомерЭл := Прямоуг (Шапка, n3, n4, n2, n1);

```
        n3 := n3 + n2 + n2 / 2;
        НомерЭл := Прямоуг (Шапка, n3, n4, n2 + n2, n1);
        n3 := n3 + n2 + n2 + n2 / 2;
        НомерЭл := Прямоуг (Шапка, n3, n4, n2, n1);
      on Вид = 2; { Вид спереди }
        НомерЭл := Прямоуг (Шапка, 0, 0, L, H);
      on Вид = 3; { Вид сбоку }
        НомерЭл := Прямоуг (Шапка, 0, 0, B, H);
    endcase;
  endprogram;
```

Откомпилированный (исполняемый) код программы включается в библиотеку ППГ, из которой удобно выбирать нужную. Исполняемый код состоит из команд, каждая из которых. записывается в виде нескольких двухбайтных элементов. Первый из них - код операции, остальные - смещения операндов относительно начала стека операции.

Перед исполнением ППГ все переменные инициализируются значениями "не определено". Значения переменных хранятся вместе с их типами. Перед выполнением любой операции производится проверка того, что все аргументы уже имеют определенные значения, и их типы допустимы для выполнения операции. Также проверяются естественные ограничения значений параметров элементов чертежа. Например, значение цвета от 0 до 15.

При исполнении ППГ изменяются глобальные установки по умолчанию для различных элементов чертежа и другие настройки. После завершения ее работы они автоматически восстанавливаются, в самой ППГ заботиться об этом не нужно. Сгенерированный чертеж автоматически предлагается поместить в нужное место, назначить цвет и др.

На рисунках 1 и 2 приведен пример использования одной из ППГ, выполняющей генерацию чертежа фундамента под оборудование.

Рис.1 Выбор и ввод параметров фундамента для генерации

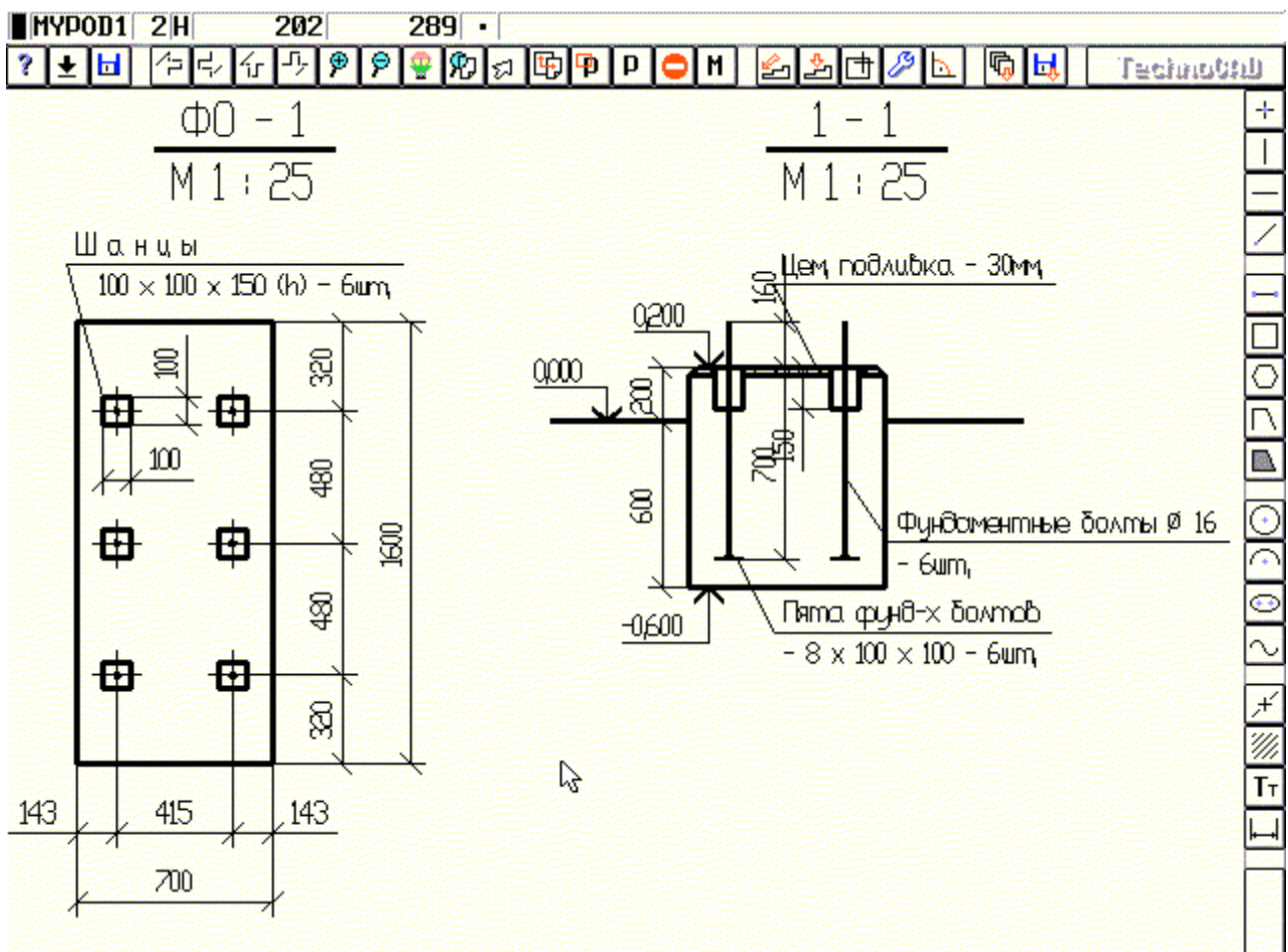

Рис.2. Сгенерированный чертеж фундамента под оборудование

Рис.3 иллюстрирует выбор библиотеки ППГ для подключения к текущему чертежу. Имеющиеся библиотеки ориентированы на потребности строительной части проекта.

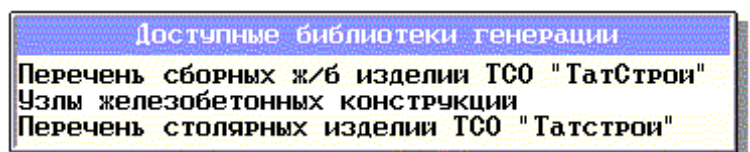

Рис.3. Выбор библиотеки ППГ для генерации чертежей

В среде разработки ППГ имеется сервис, предназначенный не только для создателя самих ППГ, но и для разработчика самой среды. Здесь унифицирована технология добавления новых встроенных операций. Для новой операции пишется одна процедура Pascal, в которой локализованы все особенности операции. Эта процедура в зависимости от текущего режима компиляции или исполнения сама выполняет все требуемые проверки, получает операнды, и выполняет операцию, выдает сообщения об ошибках. Вне этой процедуры имеются лишь сведения об имени процедуры и приоритете ее выполнения.

1. Г.Евгенев, Б.Кузьмин, С.Лебедев, Д.Тагиев. САПР XXI века: интеллектуальная автоматизация проектирования технологических процессов//САПР и графика, 2000, № 4, С.46 - 49.
2. Мигунов В.В., Потапов А.Ю., Кудрявцев Д.А., Макаров А.М., Сафин И.Т., Кафиятуллов Р.Р. Среда разработки программ параметрической генерации чертежей TechnoCAD GlassX 18 (ППГ)//Компьютерные учебные программы и инновации, 2002, № 4. – М.: Министерство образования РФ, С.45-46.